\documentstyle[preprint,aps]{revtex}

\def\q{\quad}
\def\sg{\sigma}

\newcommand{\bam}{\left( \begin{array}}
\newcommand{\eam}{\end{array} \right)}
\newcommand{\bamq}[4]{\left( \begin{array}{cccc}{#1}&{#2}&{#3}&{#4}\\}

\begin{document}
\draft
\preprint{Lecce University}
\title{Unified Octonionic Representation
of the 10-13 Dimensional Clifford Algebra
}
\author{
Khaled Abdel-Khalek\footnote{Work supported by an
ICSC--World Laboratory scholarship.\\
e-mail : khaled@le.infn.it}
}
\address{ 
Dipartimento di Fisica - Universit\`a di Lecce\\
- Lecce, 73100, Italy -}

\date{March 1997}
\maketitle
\begin{abstract}
We give a one dimensional octonionic representation of the different
Clifford algebra $Cliff(5,5)\sim Cliff(9,1),\ Cliff(6,6)\sim Cliff(10,2)
$ and lastly $Cliff(7,6)\sim Cliff(10,3)$ which can be given by
$8 \times 8$ real matrices taking into account some suitable
manipulation rules.
\end{abstract}

\widetext
\vspace{2cm}

\section{Introduction}

Since a long time, it has been conjectured that
there exists a possible connection between the different 
members of the ring division algebra (${\cal R}, {\cal C}, {\cal H},
 {\cal O}$) and the critical dimensions of the Green-Schwarz
superstring action 
\cite{kt,bc,ev}. Especially, the octonionic case has 
gained much attention due to its possible relation to the 10 dimensions
physics \cite{evan,evan1,berk1,berk2,sud,ta,mag,oda,fj,st,engl,engl1,engl2}.
Not just strings, but even extended to p-branes, octonions are usually
related to different 10 , 11 dimensions p-branes 
\cite{duff} and we would expect that the new
M, F, S theories to be no exception.

The 10 dimensions superstring (Green-Schwarz) action is
defined in terms of 16 components Majorana-Weyl spinor.
The $\kappa$ symmetry removes half of this degrees of
freedom and renders the theory dependant of just 8 real 
components. 
So, trying to write a lagrangian for
superstrings, we meet one of its many miracles, the theory
knows how to do the D=10 Lorentz group with only $8 \times 8$
matrices 
but we don't. 
In this article, we would like to show the possibility
of writing D=10 Clifford Algebra using
one single octonions. To this end, we find a surprise
the formulation allows the minimal, not just D=10 but, D=13 
Cliff(7,6).

The idea is to know how to translate some real $n\times n$ 
(${\cal R}(n)$)
 matrices to their corresponding complex and quaternionic matrices
\cite{rot1}, in general, 
which can be extended to the octonionic algebra. 
It is well known from a topological point of view that any
${\cal R}^{2n}$ is trivially a ${\cal C}^{n}$ complex 
manifold and any ${\cal R}^{4n}$ is also
a trivial quaternionic manifold
 ${\cal H}^n$, whereas, any ${\cal R}^{8n}$ is again a trivial
${\cal O}^n$ octonionic manifold. And, as any ${\cal R}^n$ is isomorphic
as a vector space to ${\cal R}(n)$ matrices, we would expect 
\begin{eqnarray}
{\cal R}(2n) \rightarrow {\cal C}(n) ; \label{cl0}\\
{\cal R}(4n) \rightarrow {\cal H}(n) ; \label{cl1}\\
{\cal R}(8n) \rightarrow {\cal O}(n) \label{cl2}.\end{eqnarray}
The aim of the paper is to illustrate how this 
structural 
isomorphism\footnote{Actually the isomorphism does not hold
for the octonionic case as it is evident
that matrix algebra is associative whereas 
octonions are not. Nevertheless, we can find
some translation rules between ${\cal O}$ and
$R^8$.} 
works and to construct 
different Clifford algebra over quaternions
and octonions. The paper is organized as follows:
In section II. , just for completion, we prove this isomorphism for
the complex numbers. In section III. we meet the new problems
due to the non-commutativity of quaternions. We construct
the quaternionic case in full details   which will be
a warm up as well as a good guide for the octonionic case
which we treated at \cite{my1} and  will be 
summariezed at section IV. emphasising the main differences and
similarities in confront the quaternionic case.
Then in section V. we show how to construct Lie
Algebra (classical) from octonions. Adding one rule to 
overcome the non-associativity, we 
 will be able to write down a Cliff(7,6) that can be represented by two sets
of $8 \times 8$ real
matrices defined with some simple manipulation rules. 
Lastly in section VI. we present
our conclusions as well as some further
open points. The construction is physically, not mathematically, motivated
since, without the strange $\kappa$ symmetry, no one would try to
look for this formulation.

\section{Complex Numbers}

To prove  
 (\ref{cl0})
, the idea goes as follows
:
For complex variables, one can represent any complex number
z as an element of ${\cal R}^2$

\begin{eqnarray}
z = z_0 + z_1 e_1,
\equiv Z = 
\left( \begin{array}{c}
z_0 \\
z_1
\end{array} \right).
\end{eqnarray}
The action of 1 and $e_1$ induce the following matrix transformations on 
Z
,
\begin{eqnarray}
1.z &=& z.1 = z \equiv Z = \openone Z ,
\end{eqnarray}
while
\begin{eqnarray} 
e_1.z &=& z.e_1 = z_0 e_1 - z_1 
\\ &\equiv& 
E_1 Z   \\
&=&
\left( \begin{array}{cc}
0 & -1 \\
1 & 0 
\end{array}
\right) 
\left(
\begin{array}{c}
z_0 \\
z_1
\end{array} \right) =
\left( \begin{array}{c}
-z_1 \\
z_0
\end{array} \right). 
\end{eqnarray}
Now, we have a problem, these two matrices $\openone$ and $E_1$
are not enough to form a basis for $R(2)$. 
The solution of our dilemma is easy. We should also
take into account 
\begin{eqnarray}
z^* = z_0 - z_1 e_1 \equiv 
Z^* = \left( \begin{array}{c}
z_0 \\
-z_1
\end{array} \right)
\end{eqnarray}
so, we find
\begin{eqnarray}
1^*.z &=& z^* 
\\ &\equiv&
\openone^* Z = Z^*\\
&=& 
 \left( \begin{array} {cc}
1 & 0 \\
0 & -1 
\end{array} \right) 
\left( \begin{array}{c}
z_0 \\
z_1 
\end{array} \right) = 
\left( 
\begin{array}{c}
z_0 \\
-z_1 
\end{array} \right) , \end{eqnarray}
and  \begin{eqnarray} 
e_1.z^* &=& z^*.e_1 =
z_0 e_1 + z_1  = e_1^* z\\
&\equiv& E_1 Z^* = E_1^* Z \\
&=& 
\left( \begin{array}{cc}
0 & -1 \\
1 & 0 
\end{array} \right) 
\left( \begin{array}{c}
z_0 \\
- z_1 
\end{array} \right) =
\left( \begin{array}{cc}
0 & 1 \\
1 & 0 
\end{array} \right) 
\left( \begin{array}{c}
z_0 \\
z_1 
\end{array} \right)
= \left( \begin{array}{c}
z_1 \\
z_0 
\end{array} \right) .
\end{eqnarray}
Having, these four matrices $\{\openone,\openone^*,E_1,E_1^*\}$, 
(\ref{cl0}) is proved.
Lastly, we would like to mention that the relation between 
 $\{\openone^*,E_1,E_1^*\}$ and the quaternionic imaginary units
, defined in the next section is the exact reason for
the possible formulation of the 2 dimensions geometry 
in terms of quaternions \cite{wood}.

\section{Quaternions}

For quaternions, being non commutative, one should differentiate 
between right and left multiplication,
(our quaternionic algebra is given by $e_i\ . e_j =  - \delta_{ij} + 
\epsilon_{ijk} e_k,
\  \mbox{and}\ i,j,k = 1,2,3)$,
\begin{eqnarray}
q &=& q_0 + q_1 e_1 + q_2 e_2 + q_3 e_3 
\equiv Q = \left( \begin{array}{c}
q_0 \\
q_1 \\
q_2 \\
q_3 \end{array} \right) 
\quad,
\label{qw} 
\end{eqnarray} then
\begin{eqnarray} 
e_1. q &=& q_0 e_1 - q_1 + q_2 e_3 - q_3 e_2  
\\ &\equiv& E_1 Q
\\ &=& 
\left( \begin{array}{cccc} 
0 & $-1$ & 0 & 0 \\
1 & 0 & 0 & 0 \\
0 & 0 & 0 & $-1$ \\
0 & 0 & 1 & 0
\end{array} \right)
\left( \begin{array}{c}
q_0 \\
q_1 \\
q_2 \\
q_3
\end{array} \right) = 
\left( \begin{array}{c}
-q_1 \\
q_0 \\
-q_3 \\
q_2 \end{array} \right) , \label{P1} \end{eqnarray}
whereas\footnote{We  use the elegant 
notations of \cite{rotl}.}
 \begin{eqnarray}
( 1|e_1 ). q &=&
q\ . e_1 = q_0 e_1 - q_1 - q_2 e_3 + q_3 e_2 
\\ 
&\equiv& 1|E_1~Q = Q E_1
\\ &=& 
\left( \begin{array}{cccc}
0 & $-1$ & 0 & 0\\
1 & 0 & 0 & 0\\
0 & 0 & 0 & 1\\
0 & 0 & $-1$ & 0
\end{array} \right)
\left( \begin{array}{c}
q_0 \\
q_1 \\
q_2 \\
q_3
\end{array} \right) = 
\left( \begin{array}{c}
-q_1 \\
q_0 \\
q_3 \\
-q_2 \end{array} \right) , \label{P2} 
\end{eqnarray}
and so on for the different ($e_2, e_3, 1|e_2, 1|e_3$)\footnote{
Notice that $\{E_1,E_2,E_3,1|E_1,1|E_2,1|E_3\}$ are the 't Hooft
matrices\cite{hoft}.}
,
\begin{eqnarray}
e_2 \equiv E_2 =
\left( \begin{array}{cccc}
0&0&-1&0\\
0&0&0&1\\
1&0&0&0\\
0&-1&0&0
\end{array} \right)
&\q,&
e_3 \equiv E_3 =
\left( \begin{array}{cccc}
0&0&0&-1\\
0&0&-1&0\\
0&1&0&0\\
1&0&0&0
\end{array} \right) ,
\label{P3} \\
1|e_2 \equiv 1|E_2 =
\left( \begin{array}{cccc}
0&0&-1&0\\
0&0&0&-1\\
1&0&0&0\\
0&1&0&0
\end{array} \right)
&\q,&
1|e_3 \equiv 1|E_3 =
\left( \begin{array}{cccc}
0&0&0&-1\\
0&0&1&0\\
0&-1&0&0\\
1&0&0&0
\end{array} \right) , \label{P4}
\end{eqnarray}
which enable us to find any generic $e_i|e_j$ 
\begin{eqnarray}
e_i|e_j.\ q  = e_i\ .\ 1|e_j\ .q = e_i\ . q .\ e_j , \label{P5}
 \end{eqnarray}
then we have  the possible 15 combinations
${\cal H}|{\cal H}$ 
\begin{equation}
\{e_1,e_2,e_3,
1|e_1,e_1|e_1,e_2|e_1,e_3|e_1,
1|e_2,e_1|e_2,e_2|e_2,e_3|e_2,
1|e_3,e_1|e_3,e_2|e_3,e_3|e_3\}.
\label{ml1}
\end{equation}
And their corresponding matrices
\begin{equation} 
\{E_1,E_2,E_3, 
1|E_1,E_1|E_1,E_2|E_1,E_3|E_1,
1|E_2,E_1|E_2,E_2|E_2,E_3|E_2,
1|E_3,E_1|E_3,E_2|E_3,E_3|E_3\}. 
\label{ml2}
\end{equation}
Using the matrices $\{E_1,E_2,E_3\}$, we
have 
($\times$ is the usual matrix multiplication), 
\begin{eqnarray}
E_i \times  E_j = -\delta_{ij} \openone + \epsilon_{ijk} E_k ,
\label{lk1}
\end{eqnarray}
they satisfy the same algebra as their corresponding
quaternionic units $\{e_1,e_2,e_3\}$ i.e they are isomorphic. 
Keep in mind this relation in order to compare it later with the octonionic
case.

We can  deduce the following  group structure
for our quaternionic operators
\begin{itemize}
\item Left $su(2)_L$ 
\begin{eqnarray}
e_i . e_j         &=& -\delta_{ij} + \epsilon_{i j k} e_k , \label{p1} 
\\su(2)_L &\sim& ~\{e_1,e_2,e_3\} . 
\end{eqnarray}
\item Right $su(2)_R$
\begin{eqnarray}
1|e_i . 1|e_j     &=&  1|(e_j.e_i) = -\delta_{ij} + 
\epsilon_{j i k} 1|e_k , \label{p2} 
\\su(2)_R &\sim& ~\{1|e_1,1|e_2,1|e_3\} . 
\end{eqnarray}
This rule can be also explicitly derived using $\{1|E_1,1|E_2,1|E_3\}$ 
.
\item $so(4) \sim su(2)_L \otimes su(2)_R$, which can
be proved using (\ref{p1}) and (\ref{p2}) and
\begin{eqnarray}
e_i . 1|e_j       &=& 1|e_j . e_i = e_i|e_j 
,
\quad \mbox{i.e} \quad [ e_i ,~1|e_j ] = 0 , \label{kl1}\\
so(4) &\sim& ~\{e_1,e_2,e_3,1|e_1,1|e_2,1|e_3\} . 
\end{eqnarray}
A weak form of (\ref{kl1}), as we will see later, holds for octonions.
\item $spin(3,2)$ - and its
subgroups - which can be proved by a Clifford Algebra 
construction
\begin{eqnarray}
&~&\gamma_0 = e_1|e_1 ,~~ 
\gamma_1 = e_3 ,~~
\gamma_2 = e_1|e_2,~~
\gamma_3 =  e_2,~~
\gamma_5 =  e_1|e_3, \label{gamma5}\\
&~&\{\gamma_\alpha , \gamma_\beta \} = 2 diag(+,-,+,-,+) .
\label{clf1}
\end{eqnarray}
Using the matrix form (\ref{P1}--\ref{P4}) and (\ref{P5}), we can calculate
these $\gamma$'s, we find
\begin{eqnarray}
\gamma_0 = 
\left( \begin{array}{cccc}
-1&0&0&0\\
0&-1&0&0\\
0&0&1&0\\
0&0&0&1
\end{array} \right)
&\q& 
\gamma_1 = 
\left( \begin{array}{cccc}
0&0&0&-1\\
0&0&-1&0\\
0&1&0&0\\
1&0&0&0
\end{array} \right)
\\
\gamma_2 = 
\left( \begin{array}{cccc}
0&0&0&1\\
0&0&-1&0\\
0&-1&0&0\\
1&0&0&0
\end{array} \right)
&\q& 
\gamma_3 = 
\left( \begin{array}{cccc}
0&0&-1&0\\
0&0&0&1\\
1&0&0&0\\
0&-1&0&0
\end{array} \right)
\\
~~~~~~~~~~~~~\gamma_5 = 
\left( \begin{array}{cccc}
0&0&-1&0\\
0&0&0&-1\\
-1&0&0&0\\
0&-1&0&0
\end{array} \right).
\end{eqnarray}
It is clear that these matrices are
are nothing but the famous Dirac representation up to a minus sign 
and the standard $\gamma_2$ is multiplied by $-i$.
By explicit calculation, one finds (in the basis given above)
\begin{eqnarray}
spin(3,2) &\sim&
~\{[ \gamma_\alpha , \gamma_\beta ] \} \quad \quad
\alpha , \beta = 0,1,2,3,5 \quad , \\
&\sim&~\{e_1,1|e_1,1|e_2,1|e_3,
e_2|e_1,e_3|e_1,e_2|e_2,e_3|e_2,e_2|e_3,e_3|e_3\} .
\end{eqnarray}
Actually, the main reason for this $spin(3,2)$ is
the following relation 
\begin{eqnarray}
e_i .e_j. ~1|e_k + e_j .e_i .~1|e_k = 0 \label{tl1} ,
\end{eqnarray}
this construction is well known since
a long time and used by Synge \cite{Synge} to give a quaternionic formulation
of special relativity ($so(1,3)$) but we don't know who
was the first to derive it (most probable is Conway but the
reference is too old and rare to find). Just we have done it here explicitly
in parallel with their corresponding $4\times 4$ real matrices and avoided
the use of complexified quaternions.
\item  
Also at the matrix level the full set ${\cal H}|{\cal H}$ closes an algebra,
\begin{eqnarray}
1|e_i . e_j|e_k   &=& \epsilon_{kil} e_j|e_l ,\label{il3}\\
e_i . e_j|e_k     &=& e_j|e_k . e_i = \epsilon_{i j l} e_l|e_k ,\label{il4}\\
e_i|e_j . e_m|e_n &=& \epsilon_{i m l} \epsilon_{n j p} e_l|e_p \label{il5}. 
\end{eqnarray}
By explicit calculations, we found that it is impossible
to construct a sixth $\gamma$ from the
 set ${\cal H}|{\cal H}$. The easiest way to see
this, is to use (\ref{gamma5}) and calculate
\begin{eqnarray}
\gamma_0.\gamma_1.\gamma_2.\gamma_3.\gamma_5 = -1 ,
\end{eqnarray}
so it does not form  any $spin(n,m)$ higher than 
$spin(3,2)$.
\item 
Adding the identity to ${\cal H}|{\cal H}$, we used Mathematica
to prove that these 16 matrices are linearly independent
so they can form a basis for any $R(4)$ as we claimed 
in (\ref{cl1}). 
\end{itemize}

A big difference between octonions and quaternions is the following
: All the last equations can be reproduced by matrices {\bf exactly} by
replacing $e \longrightarrow E$ i.e
there is an isomorphism between
(\ref{ml1}) and (\ref{ml2}). The isomorphism
can be derived explicitly between (\ref{lk1})
 and (\ref{p1}) ,then by deriving the suitable rules at the quaternionic
level (\ref{p2},\ref{kl1},\ref{il3},\ref{il4},\ref{il5}),
 it can be extended to the whole set of left and right actions
as well as their mixing. 
In the octonionic case only the 
Clifford algebraic construction resists and holds.

\section{Octonions}

Moving to octonions, we  use the  
symbols $e_i$ to denote the imaginary octonionic units
where $i,j,k = 1 .. 7$ and 
\begin{eqnarray}
e_i\ . e_j = -\delta_{ij} + 
\epsilon_{i j k} e_k \quad \mbox{or} \quad
[e_i,e_j] = 2 \epsilon_{ijk} e_k ,
\label{oiu}
\end{eqnarray}
such that $\epsilon_{i j k}$ equals 1 for one of the following 
seven combinations \{(123), (145), (176), (246), (257), (347), (365)\} ,
also, we  use the symbol g to represent a generic octonionic
number,  and 
its corresponding element over ${\cal R}^8$ is 
denoted by G.
As octonions are non-associative, we meet new surprises. 
We would like to
sketch the main differences between the octonionic and
the quaternionic case outlining briefly the construction given in 
\cite{my1,schaf} and focusing on     
the  Clifford algebraic structure : 
The translation rules 
go exactly in the same manner as (\ref{qw}-\ref{P4}), just
we have to use this time $R(8)$ instead of $R(4)$, we find 
\begin{itemize}
\item First:
Our left and right matrices are no more isomorphic to the octonionic
algebra, for left action, 
we have
\begin{eqnarray}
[E_i,E_j] = 2 \epsilon_{ijk} E_k - 2 [E_i,1|E_j] ,
\end{eqnarray}
while
\begin{eqnarray}
[e_i,e_j] = 2 \epsilon_{ijk} e_k,
\end{eqnarray}
so the isomorphism at the level of algebra is lost and actually
can never be restored as  matrices
are associative but octonions are not. Moreover
the set $\{E_i\}$ alone does not close an algebra. Including
the right action in our treatment is an obligation not a choice, 
 then, we will
be able to find something useful as we will see.

For right action, the situation is the following
\begin{eqnarray}
[1|E_i,1|E_j] = 2 \epsilon_{jik}1|E_k - 2[E_i,1|E_j] ,
\end{eqnarray}
and we have
\begin{eqnarray}
[1|e_i,1|e_j] = 2 \epsilon_{jik} ~1|e_k .
\end{eqnarray}
\item Second: The anticommutation relations hold  at 
the octonionic and matrix level
\begin{eqnarray}
~\{e_i,e_j\}~=~\{1|e_i,1|e_j\}~= -2 \delta_{ij} ,
\label{clif0}
\end{eqnarray}
and  the same for $E_i$ and $1|E_i$ ,
\begin{eqnarray}
~\{E_i,E_j\}~=~\{1|E_i,1|E_j\}~= -2 \delta_{ij} \openone.
\label{clif1}
\end{eqnarray}
 So a Clifford algebraic construction will be possible.
\item Third: Due to the non-associativity, 
\begin{eqnarray}
(e_1.(e_2 .g)) \neq ((e_1.e_2).g) ,
\end{eqnarray}
we have to introduce left/right
octonionic operators, 
\begin{eqnarray}
e_i(e_j\ . g = e_i\ .\ (g. e_j ) \equiv R_{ij} \times G ,\\
e_i)e_j\ . g = (e_i. g)\ .\ e_j \equiv L_{ij}  \times G ,
\end{eqnarray}
which can be constructed from the following sets,
$\{ e_1,\ ...\ , e_7, 1|e_1,\ ... \ , 1|e_7\}$ and
$\{ E_1,\ ...\ , E_7, 1|E_1,\ ... \ , 1|E_7\}$, 
as follows
\begin{eqnarray}
e_i(e_j.g = e_i.1|e_j.g \equiv R_{ij} = E_i
\times 1|E_j \times G ,\\
e_i)e_j.g = 1|e_j.e_i.g \equiv L_{ij} = 1|E_j \times E_i \times G .
\end{eqnarray}
We have given the matrix form of
$\{ E_1,\ ...\ , E_7, 1|E_1,\ ... \ , 1|E_7\}$ 
in a separate appendix.
Having this set, we can form the different 106 elements
\begin{eqnarray}
\begin{array}{cc}
1, ~ e_{m}, ~ 1\mid e_{m} &  ~~~~~(15 elements) ,\\
e_{m}\mid e_{m} & (7) ,\\
e_{m}~)~e_{n}~~~~{(m\neq n)} &  (42) ,\\
e_{m}~(~e_{n}~~~~{(m\neq n)} &  (42) ,\\
{(m, \; n =1, \dots, 7) \q .} & 
\end{array}
\end{eqnarray}

but the two sets of 42 left/right operators are actually
linearly dependent , so we should constrain ourselves to either 
one of them leaving for us just 64 elements.
Thus, the use of this left/right operators
is necessary to overcome the non-associativity problem 
and, at the same time,  to form a basis 
of $R(8)$. 
\end{itemize}

\section{The Clifford Algebraic Construction}

The easiest way to construct a Lie algebra from
our left/right octonionic operator is to use a
Clifford algebraic construction.
As it is clear from  (\ref{clif0}), any of the 
set $\{e_i\}$ or $\{1|e_i\}$ gives an octonionic representation
of Cliff(0,7) which can be represented by the matrices
$\{E_i\}$ or $\{1|E_i\}$ allowing us to construct, for example, 
the following spin algebra : 
\begin{itemize}
\item Matrix representation of $so(7)_L$
\begin{eqnarray}
so(7) \sim \{ [E_i,E_j] \} \quad \quad i,j = 1...7
\end{eqnarray}
\item Matrix representation of $so(7)_R$
\begin{eqnarray}
so(7) \sim \{ [1|E_i,1|E_j] \} \quad \quad i,j = 1...7
\end{eqnarray}
\item Matrix representation of $so(8)_L$
\begin{eqnarray}
so(8) \sim S_7\otimes so(7) \sim 
\{E_i, [E_i,E_j] \} \quad \quad i,j = 1...7
\end{eqnarray}
\item Matrix representation of $so(8)_R$
\begin{eqnarray}
so(8) \sim S_7 \otimes so(7) \sim
\{ 1|E_i, [1|E_i,1|E_j] \} \quad \quad i,j = 1...7
\end{eqnarray}
where $S_7$ is the Reimannian seven sphere.
\end{itemize}
Notice that, from (\ref{oiu}), we can not use 
the $e_i$ as they are non-associative and consequently
they don't satisfy the Jacobi identity.
In summary, whatever
our left/right matrices do not form an isomorphic
representation of our left/right octonionic
operators, they admit an isomorphic Clifford algebra.
Now, trying to have something larger than
Cliff(0,7) like the 
 quaternionic Cliff(3,2) (eqn. ~\ref{clf1}),
one would try
\begin{eqnarray}
\gamma_0 &\rightarrow& e_2 , \quad
\gamma_1 \rightarrow e_3 ,\quad
\gamma_2 \rightarrow e_4 ,\nonumber \\
\gamma_3 &\rightarrow& e_5 ,\quad
\gamma_4 \rightarrow e_6 ,\quad
\gamma_5 \rightarrow e_7 ,\nonumber \\
\gamma_6 &\rightarrow& e_1(e_1 ,\quad
\gamma_7 \rightarrow e_1(e_2 ,\quad
\gamma_8 \rightarrow e_1(e_3 ,\nonumber \\
\gamma_9 &\rightarrow& e_1(e_4 ,\quad
\gamma_{10} \rightarrow e_1(e_5 ,\quad
\gamma_{11} \rightarrow e_1(e_6 ,\nonumber \\
\gamma_{13} &\rightarrow& e_1(e_7. 
\label{kliff}
\end{eqnarray}

But this construction works well for $\gamma_{0..5}$ but fails
elsewhere, for example
\begin{eqnarray}
\{ \gamma_0,\gamma_1 \} g &=&
e_2 ( e_3 ( g_0 e_0 +
g_1 e_1 + g_2 e_2 + g_3 e_3 + g_4 e_4 + g_5 e_5 
+ g_6 e_6 + g_7 e_7 ) )
\nonumber \\ 
&+& 
e_3 ( e_2 ( g_0 e_0 +
g_1 e_1 + g_2 e_2 + g_3 e_3 + g_4 e_4 + g_5 e_5 
+ g_6 e_6 + g_7 e_7 ) )
\nonumber \\
&=& 0.
\end{eqnarray}
whereas
\begin{eqnarray}
\{ \gamma_0 , \gamma_8 
\} &=& 
e_2 ( e_1 ( ( g_0 e_0 +
g_1 e_1 + g_2 e_2 + g_3 e_3 + g_4 e_4 + g_5 e_5 
+ g_6 e_6 + g_7 e_7 ) e_3 ) )
\nonumber \\
&+&
e_1 ( ( e_2  ( g_0 e_0 +
g_1 e_1 + g_2 e_2 + g_3 e_3 + g_4 e_4 + g_5 e_5 
+ g_6 e_6 + g_7 e_7 ) ) e_3 )
\nonumber \\
&\neq& 0. \label{nml1}
\end{eqnarray}
One may give up and say octonions are different from 
quaternions and they are non-associative. But, because of
this reason, we still have more freedom.
By a careful analysis of (\ref{nml1}), it becomes clear
that the reason of the failure is
\begin{eqnarray}
E_i \times ~1|E_j \neq 1|E_j \times E_i
\end{eqnarray} 
a weaker form holds
\begin{eqnarray}
E_i \times ~1|E_i = 1|E_i \times E_i
\end{eqnarray} 
in complete contrast with (\ref{kl1}). The solution can
be found to get around this problem.
 
Because of the non-associativity, we should give to
left and right action different priorities.
As a matter of fact, this is a very reasonable requirement.
When we transferred from complex numbers to quaternions,
we introduced barred operators in order to overcome the
non-commutativity problem and we defined their consistent
rules, so going to octonions, we should expect more rules.

Assuming higher priority to right action i.e
\begin{eqnarray}
e_1(e_2\ .\ e_4\ .\ g \equiv (e_1.(e_4.(g.e_2))) , \\
e_4\ .\ e_1(e_2\ .\ g \equiv  (e_4.(e_1.(g.e_2))) .
\end{eqnarray}
then
\begin{eqnarray}
\{\ e_1(e_2\ ,\ e_4 \ \}. g = 0.
\end{eqnarray}
Using these simple rules, we can generalize
 (\ref{gamma5}). 
Using the following identities
\begin{eqnarray}
\{ E_i, E_j \} = -2 \delta_{ij} \openone,\\
\{ 1|E_i, 1|E_j \} = -2 \delta_{ij} \openone ,\\
E_i \times E_j \times 1|E_k + E_j \times E_i \times 1|E_k 
= 0 ,
\end{eqnarray}
which hold equally well at the octonionic level
\begin{eqnarray}
\{ e_i, e_j \} = -2 \delta_{ij} ,\\
\{ 1|e_i, 1|e_j \} = -2 \delta_{ij} ,\\
e_i . e_j . 1|e_k + e_j . e_i . 1|e_k 
= 0 ,
\end{eqnarray}
in complete analogy with (\ref{tl1}).
Now, we have the possibility to write down
the $Cliff(7,6)$ which is given in (\ref{kliff}).

When any of the $\gamma_{6..13}$'s are translated into matrices,
each one has  two different forms, depends of being
acted from right or left, e.g.
\begin{eqnarray}
\gamma_0 \gamma_9 = 
e_2.e_1(e_4.g \equiv E_2 \times E_1 \times 1|E_4 \times G ,
\label{lp1}\\
\gamma_9 \gamma_0 = 
e_1(e_4.e_2.g \equiv E_1 \times E_2 \times 1|E_4 \times G,
\label{lp2}
\end{eqnarray} 
{\bf they don't have a faithful $8\times 8$ matrix representation}.
To be clear, in (\ref{lp1}), we say that $\gamma_9$ is represented
by the matrix $E_1 \times ~1|E_4$ but in (\ref{lp2}) this 
statement is not valid anymore as $E_2$ is now sandwiched
between the $E_1$ and $1|E_4$.
This is a very important fact and should be always taken into
account. 
When we count the numbers of degrees of freedom,
 we have 64 for left action and 64 for right action,
 in total 128 real parameters which are enough
to represent our $Cliff(7,6)$. 

Actually, because octonions are non-associative, sometimes, 
we can do
with them what we can not do with matrices in a straightforward
way.

\section{conclusion and Further Developments}

As it happened many times, physics may lead mathematics.
We had a problem in the start. Guided by quaternions,
and using octonions, we found a simple way to do $Cliff(7,6)$
by $8 \times 8$ matrices. One would, even from the start, 
work with matrices but we think it wouldn't be obvious at all
that our $\gamma$'s are represented by two different matrices
depending upon left or right action.

Apart from the $S_7$ compactification of the D=11 supergravity
and the exceptional Gut groups, the application of
octonions in physics may be called ``The Art of Conjectures''.
As they are non-associative, it is widely believed that they
don't admit any Hilbert space construction. Fortunately,
this is completely wrong as had been shown in \cite{bh},
where many of the  machinery of functional analysis had
been given for  this Octonionic Hilbert space by imposing
different forms of scalar product (also see \cite{remb}).
Mainly, such construction uses a $Cliff(0,7)$ instead
of using $S_7$ \footnote{The idea is
somehow similar to the construction of the index theorems
 where one uses the Clifford bundle instead of
the de Rham bundle of exterior differential forms \cite{lawson}.}.
Such techniques had been applied successfully
at the quaternionic level where the complex scalar
product plays a fundamental role. A complete
formulation of the standard model had been given in \cite{rot2} 
 and we had extended such
methods to the octonionic level \cite{my2}.

Finally, we want to comment about the possible further
applications and investigations:

1- The Green-Schwarz string action in $D=10$
 depends on a 16-real components Majoranna-Weyl
spinor, the $\kappa$ symmetry removes half of these fermionic 
degrees of freedom leaving the action depends on just
8 real fermionic components i.e one octonion \cite{ced}.
Since, there is {\bf no way} to find 
D=10 dimensions Clifford algebra $8\times 8$ gamma matrices,
this represents an obstacle towards a covariant string formulation.
Our representation is dependent on exactly one octonion i.e 8 real components.
Actually, this was the main motive of this work. Superstring exists
and without any doubt it is our best candidate for the dreamed theory
of every thing, finding its true formulation is highly required.
 Can it be the octonionic string \cite{st}!

2- The unified 10-13 dimensions octonionic 
representation is in agreement with the recent discovery of 13
hidden dimensions in string theory \cite{bar}.
 It would be easier to
work with one octonionic construction
instead of 32 components gamma matrices.

3- What is the real meaning of the different p-branes dualities?
May be nothing but a non-trivial mapping between
their different -- postulated -- infinite dimensional world-volume symmetries.
Or more attractively, different mapping between
different infinite dimensional ring-division superconformal algebra
which may be the real connection between the ring-division algebra and the 
p-brane program. One of the simple formula that 
holds for many p-branes is
\begin{eqnarray}
D - p = 2^n \quad \quad n =  0, 1, 2, 3.
\end{eqnarray}
Does it really mean that any consistent p-brane
should enjoy a superconformal algebra on its transverse
dimensions?
This can be an amplified form of our old problem, what is the
correct relation
 between the string sheet and the
target space formulation of string theory?

We understand that the approach discussed here 
may not be the best in the market but with our potentially need
of developing and examining the recent string dualities, it seems
worthwhile to try every possible avenue.

\acknowledgments

I would like to acknowledge 
P. Rotelli and S. De~Leo as well as the physics department
at Lecce university for their kind hospitality.
Also, I am grateful to Prof. A.~Zichichi 
and the ICSC--World Laboratory for financial
support.

\newpage
\appendix
\section*{}

We introduce the following notation:

\begin{eqnarray}
\begin{array}{ccccc}
\{a,b,c,d\}_{(1)} ~ &\equiv& ~\bamq{a}{0}{0}{0} 0 & b & 0 & 0\\
0 & 0 & c & 0\\ 0 & 0 & 0 & d \eam    \q ,  
\{a,b,c,d\}_{(2)} ~ &\equiv& ~\bamq{0}{a}{0}{0} b & 0 & 0 & 0\\
0 & 0 & 0 & c\\ 0 & 0 & d & 0 \eam \q ,\\
\{a,b,c,d\}_{(3)} ~ &\equiv& ~\bamq{0}{0}{a}{0} 0 & 0 & 0 & b\\
c & 0 & 0 & 0\\ 0 & d & 0 & 0 \eam    \q , 
\{a,b,c,d\}_{(4)} ~ &\equiv& ~\bamq{0}{0}{0}{a} 0 & 0 & b & 0\\
0 & c & 0 & 0\\ d & 0 & 0 & 0 \eam \q ,
\end{array}
\end{eqnarray}
where $a, \; b, \; c, \; d$ and $0$ represent $2\times 2$ real matrices.

In the following  $\sg_{1}$, $\sg_{2}$, $\sg_{3}$ represent the 
standard Pauli matrices.

\begin{eqnarray}
\begin{array}{ccccccc}
E_{1}  &=&\{-i\sg_{2}, -i\sg_{2}, -i\sg_{2},  
i\sg_{2} ~\}_{(1)} \q &,&
1\mid E_{1} &=& \{-i\sg_{2},  i\sg_{2}, i\sg_{2},  
-i\sg_{2} ~\}_{(1)}\q , \\
E_{2}  &=&\{ -\sg_{3}, \sg_{3}, -1, 1 ~\}_{(2)}\q &,&
1\mid E_{2} &=& \{ -1, 1, 1,  
-1 ~\}_{(2)}\q , \\
E_{3} &=&\{ -\sg_{1}, \sg_{1}, -i\sg_{2},  
-i\sg_{2} ~\}_{(2)}\q &,&
1\mid E_{3} &=& \{ -i\sg_{2}, -i\sg_{2}, i\sg_{2},  
i\sg_{2} ~\}_{(2)}\q , \\
E_{4} &=&\{ -\sg_{3}, 1, \sg_{3}, -1 ~\}_{(3)}\q &,&
1\mid E_{4} &=& \{ -1, -1, 1,  
1 ~\}_{(3)}\q , \\
E_{5} &=&\{ -\sg_{1}, i\sg_{2}, \sg_{1},  
i\sg_{2} ~\}_{(3)}\q &,&
1\mid E_{5} &=& \{ -i\sg_{2}, -i\sg_{2}, 
-i\sg_{2},  
-i\sg_{2} ~\}_{(3)}\q , \\
E_{6} &=&\{ -1, -\sg_{3}, \sg_{3}, 1 ~\}_{(4)}\q &,&
1\mid E_{6} &=& \{ -\sg_{3}, \sg_{3}, -\sg_{3},  
\sg_{3} ~\}_{(4)}\q , \\
E_{7} &=&\{ -i\sg_{2}, -\sg_{1}, \sg_{1},  
-i\sg_{2} ~\}_{(4)}\q &,&
1\mid E_{7} &=& \{ -\sg_{1}, \sg_{1}, -\sg_{1},  
\sg_{1} ~\}_{(4)}\q .
\end{array}
\end{eqnarray}
By explicit calculation, \cite{ok}, one finds that these matrices
are the direct generalization of the 't Hooft matrices \cite{hoft},
for $a = 1 .. 7$,
\begin{eqnarray}
&~&(E_a)_{\mu \nu} = \epsilon_{a \mu \nu}~~~~~~~~~~~~~~\q \q 
\mu,\nu = 1.. 7\\
&~&(1|E_a)_{\mu \nu} = - \epsilon_{a \mu \nu}~~~~~~~~~~\q \q 
\mu, \nu = 1 .. 7\\
&~&(E_a)_{0\nu} = (1|E_a)_{0\nu} = - \delta_{a\nu}~\q \q 
\mu, \nu = 0 .. 7\\
&~&(E_a)_{\mu0} = (1|E_a)_{\mu0} =  \delta_{a\mu}\q \q 
\mu, \nu = 0 .. 7\\
&~&(E_a)_{00} = (1|E_a)_{00} = 0.
\end{eqnarray}

\end{document}